\documentclass[a4,referee,pre,showpacs]{revtex4}

\usepackage[dvips]{graphicx}
\usepackage{amsmath}

\def\EQ{\begin{equation}}
\def\EN{\end{equation}}
\def\EQA{\begin{eqnarray}}
\def\ENA{\end{eqnarray}}
\def\uu{{\bf u}}
\def\vv{{\bf v}}

\def\UU{{\bf U}}

\def\A{\mathcal{A}}

\def\OOmega{{\mbox{\boldmath$\Omega$}}}
\def\OB{\bar{\Omega}}
\def\OBN{\Omega_0}
\def\ie{{\em i.e. }}
\def\T{\mathcal{T}}
\def\L{\mathcal{L}}

\begin{document}
\title{Analytical theory of forced rotating sheared turbulence. II. Parallel case}

\author{Nicolas Leprovost and Eun-jin Kim}
\affiliation{Department of Applied Mathematics, University of Sheffield, Sheffield S3 7RH, UK}

\begin{abstract}
Forced turbulence combined with the effect of rotation and shear flow is studied. In a previous paper [Leprovost and Kim, PRE in press (2008)], we considered the case where the shear and the rotation are perpendicular. Here, we consider the complementary case of parallel rotation and shear,  elucidating how rotation and flow shear influence the generation of shear flow (e.g. the direction of energy cascade), turbulence level,  transport of particles and momentum. We show that turbulence amplitude and transport are always quenched due to strong shear ($\xi =\nu k_y^2 / \A \ll 1$, where $\A$ is the shearing rate, $\nu$ is the molecular viscosity and $k_y$ is a characteristic wave-number of small-scale turbulence), with stronger reduction in the direction of the shear than those in the perpendicular directions. In contrast with the case where rotation and shear are perpendicular, we found that rotation affects turbulence amplitude only for very rapid rotation ($\Omega \gg \A$) where it reduces slightly the anisotropy due to shear flow. Also, concerning the transport properties of turbulence, we find that rotation affects only the transport of particle and only for rapid rotation, leading to an almost isotropic transport (whereas, in the case of perpendicular rotation and shear, rotation favors isotropic transport even for slow rotation). Furthermore, the interaction between the shear and the rotation is shown to give rise to non-diffusive flux of angular momentum ($\Lambda$-effect), even in the absence of external sources of anisotropy, which can provide a mechanism for the creation of shearing structures in astrophysical and geophysical systems. 
\end{abstract}


\maketitle

\section{Introduction}
\begin{figure}
\begin{center}
\includegraphics[scale=0.7,clip]{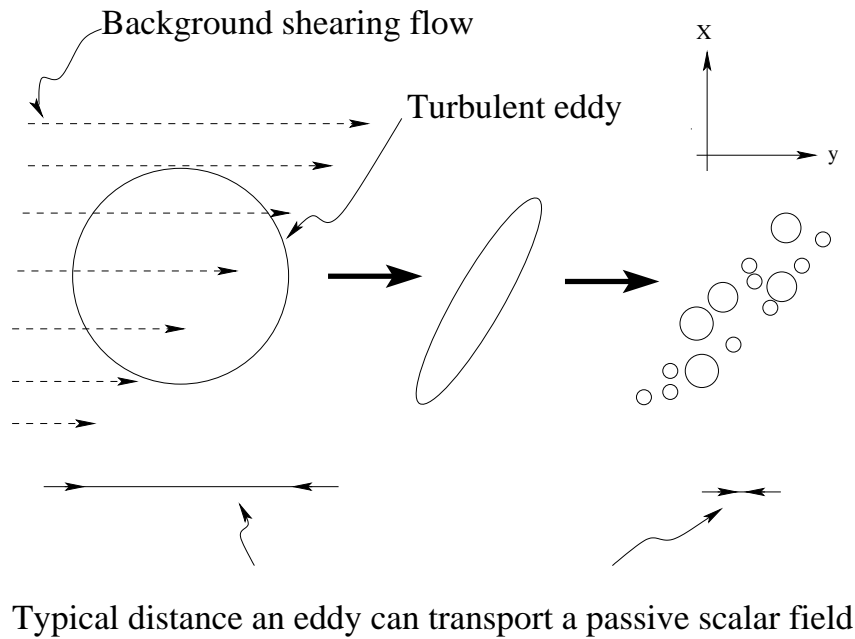}
\caption{\label{ShearEff} Sketch of the effect of shear on a turbulent eddy.}
\end{center}
\end{figure}

Large-scale shear flows are often observed in rotating astrophysical and geophysical systems. Shear and rotation have a huge impact on the properties of this system, such as energy transfer or mixing. From a physical point of view, the main effect of shear flow is to reduce turbulence level as well as turbulent transport compared to their values without shear. This is basically because shear advects turbulent eddies differentially, elongating and distorting their shapes, thereby rapidly generating small scales which are ultimately disrupted by molecular dissipation on small scales (see Fig. \ref{ShearEff}). That is, flow shear facilitates the cascade of various quantities such as energy or mean square scalar density to small scales (i.e. direct cascade) in the system, enhancing their dissipation rate. As a result, turbulence level as well as turbulent transport of these quantities can be significantly reduced compared to the case without shear. Another important consequence of shearing is to induce anisotropic transport and turbulent level since flow shear directly influences the component parallel to itself (i.e. $x$ component in Fig. 1) via elongation while only indirectly the other two components (i.e. $y$ and $z$ components in Fig. 1) through enhanced dissipation. Rotation can also reduce transport in the limit of rapid rotation (similarly to flow shear), but through a physical mechanism that is different from that of shear, namely by phase mixing of inertial waves \cite{Cally91}. While both rotation and (stable) shear flow tend to regulate turbulence, there are important differences in their effects, which should be emphasized. Rotation, by exciting inertial waves, tends to reduce turbulence transport more heavily than turbulence amplitude while shear flows reduce both of them to a similar degree. That is, rotation (or waves) quenches the cross-phase (normalized flux) more than shear flow does \cite{Kim03b,Kim06}. 

Rapid distortion theory (RDT) \cite{Batchelor54,Townsend76} was used to study the linear response of turbulence to a mean flow with spatially uniform gradients. The linear treatment of fluctuations by incorporating strong flow shear was also used in the astrophysical context by \cite{Goldreich64} by using shearing coordinates. The generation of large-scale shear flows (the so-called zonal flows) through a similar nonlocal interaction has been intensely studied in the magnetically confined plasmas, where turbulence quenching by shear flow is believed to be one of the most promising mechanisms for improving plasma confinement \cite{Burrell97,Kim04a}.  In decaying sheared turbulence, \cite{Lee90} have shown a surprisingly good agreement between the RDT predictions and numerical simulations. Subsequently, theoretical predictions (using a quasi-linear theory) for the transport of passive scalar fields in 2D hydrodynamic turbulence by \cite{Kim03b} and \cite{Kim04b} have been beautifully confirmed by recent numerical simulations \cite{Leconte06}. In particular, they have shown that turbulent transport of particles can be severely quenched inversely proportional to flow shear $\A$ while turbulence level is reduced as $\A^{-5/3}$. It is important to note that this nonlocal interaction leading to inverse cascade can be successfully captured by inhomogeneous RDT theory which permits the feedback of the nonlinear local interaction between small scales onto the large scales via Reynolds stress (constituting the other part of quasi-linear analysis) while neglecting nonlinear local interaction between small scales for fluctuations compared to nonlocal interactions. As must be obvious by comparing the Coriolis force with nonlinear advection terms, the RDT works well for sufficiently strong rotation (small Rossby number) even in the absence of shear flow. For instance, the agreement of the RDT prediction with numerical results has been shown by various previous authors including \cite{Cambon97}, but mostly in decaying turbulence. However, in this case, the RDT cannot accurately capture the turbulence structure in the plane perpendicular to rotation axis where nonlinear local interactions between inertial waves seem important (see, e.g. \cite{Smith99}). The validity and weakness of the RDT together with comparison with various numerical simulation (without an external forcing) with/without shear flows and stratification can be found in excellent review by \cite{Salhi06} and Cambon and \cite{Salhi07b}, to which readers are referred for more details. 

In  comparison, far much less is understood in the case of forced turbulence. In particular, the main interest in forced turbulence is a long-term behavior where the dissipation, enhanced by shear distortion, is balanced by energy input, thereby playing a crucial role in leading to a steady equilibrium state. The computational study of this long time behavior is however not only expensive but also difficult because of the limit on numerical accuracy, as noted by \cite{Salhi97}. Therefore, analytical theory by capturing shearing effect (such as quasi-linear theory with time-dependent wavenumber) would be extremely useful in obtaining physical insights into the problem as well as guiding future computational investigations. We note that the previous works by Kichatinov and Rudiger and collaborators \cite{Rudiger80,Kichatinov86b,Kichatinov87,Rudiger89,Kichatinov94} using quasi-linear theory are valid only in the limit of weak shear. Forced sheared turbulence was proposed for the first time by \cite{Nazarenko00a} in the context of two-dimensional near-wall turbulence to explain the logarithmic dependence of the large scale velocity on the distance to the wall. In that case, the external forcing is provided by a continuous supply of vorticity from intermittent coherent burst of vorticity coming from the viscous layer. This work was later generalized to three dimensions \cite{Nazarenko00b,Dubrulle01} with the same conclusions. In the astrophysical context, \cite{Kim05} has shown that in 3D forced HD turbulence, strong flow shear can quench turbulence level and transport of particles with strong anisotropy (much weaker along the flow shear which is directly affected by shearing) and has emphasized the difference in turbulence level and transport, which is often used interchangeably in literature. A similar weak anisotropic transport was shown for momentum transport by \cite{2Shears} in forced 3D HD turbulence. Further investigations have been performed on turbulent transport in forced turbulence by incorporating the interaction of sheared turbulence with different types of waves that can be excited due to magnetic fields \cite{Kim01,Kim06,BetaPlane}, stratification \cite{Stratification} or both magnetic fields and stratification \cite{Dynamics}. 

The combined influence of shear flow and rotation in forced turbulence has been considered by Leprovost and Kim in paper I \cite{Perpendicular08} when the rotation and the shear are perpendicular to each other. We found that flow shear always leads to weak turbulence with an effectively stronger turbulence in the plane perpendicular to shear than in the shear direction, regardless of rotation rate. The anisotropy in turbulence amplitude is however weaker in the rapid rotation limit ($\Omega \gg \A$ where $\Omega$ and $\A$ are the rotation and shearing rate) than that in the weak rotation limit ($\Omega \ll \A$) since rotation favors almost-isotropic turbulence. Compared to turbulence amplitude, particle transport is found to crucially depend on whether rotation is stronger or weaker than flow shear. When rotation is stronger than flow shear, the transport is inhibited by inertial waves, being quenched inversely proportional to the rotation rate while in the opposite case, it is reduced by shearing. Furthermore, the anisotropy is found to be very weak in the strong rotation limit (by a factor of 2) while significant in the strong shear limit. The turbulent viscosity is found to be negative with inverse cascade of energy as long as rotation is sufficiently strong compared to flow shear while positive in the opposite limit of weak rotation . Even if the eddy viscosity is negative for strong rotation, flow shear, which transfers energy to small scales, has an interesting effect by slowing down the rate of inverse cascade with the value of negative eddy viscosity decreasing as $|\nu_T| \propto \A^{-2}$ for strong shear. Furthermore, the interaction between the shear and the rotation is shown to give rise to a novel non-diffusive flux of angular momentum known as the anisotropic kinetic $\alpha$-effect (AKA) \cite{Frisch87} or as the $\Lambda$-effect in the astrophysical community. The appearance of non-diffusive term in the transport of angular momentum prevents a solid body rotation from being a solution of the Reynolds equation \cite{Lebedinsky41,Kippenhahn63}, and thus act as a source for the generation of large-scale shear flows. For instance, this effect has been advocated as a robust mechanism to explain the differential rotation in the solar convective zone. Starting from Navier-Stokes equation, it is possible to show that these fluxes arise when there is a cause of anisotropy in the system, either due to an anisotropic background turbulence (see \cite{Rudiger89} and references therein) or else due to inhomogeneities such as an underlying stratification. In \cite{Perpendicular08}, we found that a $\Lambda$-effect appears in sheared-rotating turbulence even in the absence of external sources of anisotropy. This is because the shear induces an anisotropic turbulence which combined to the rotation gives rise to non-diffusive fluxes.

In this paper, we consider the complementary case when rotation and shear are parallel to each other. By assuming either sufficiently strong shear or rotation rate, we employ a quasi-linear analysis to compute turbulence level, eddy viscosity, and particle transport for temporally short-correlated, homogeneous forcing. As the computation of these quantities involve too complex integrals to be analytically tractable, they are analytically computed by assuming an ordering in time scales. In our problem, there are three important (inverse) time-scales: the shearing rate $\A$, the rotation rate $\Omega$ and the diffusion rate $\mathcal{D}=\nu k_y^2$ where $\nu$ is the (molecular) viscosity of the fluid and $k_y^{-1}$ is a characteristic small scale of the system. We first distinguish the two cases of strong rotation ($\Omega \gg \A$) and weak rotation ($\Omega \ll \A$). The first regime of strong rotation will be studied in the strong shear ($\A \gg \mathcal{D}$) and weak shear ($\A \ll \mathcal{D}$) regime. On the other hand, the second regime of weak rotation will be considered only in the strong shear ($\A \gg \mathcal{D}$) case, as the effects of both shear and rotation disappear in the opposite limit ($\A \ll \mathcal{D}$). 

\section{Model}
\label{Model}

We consider an incompressible fluid  in a rotating frame with average rotation rate $\tilde{\Omega}$, which are governed by
\EQA
\label{NSrotation}
\partial_t \uu + \uu \cdot \nabla \uu &=& - \nabla P + \nu \nabla^2 \uu + {\bf F} - 2 \tilde{\OOmega} \times \uu \; , \\ \nonumber
\nabla \cdot \uu &=& 0 \; .
\ENA

\begin{figure}
\begin{center}
\includegraphics[scale=1,clip]{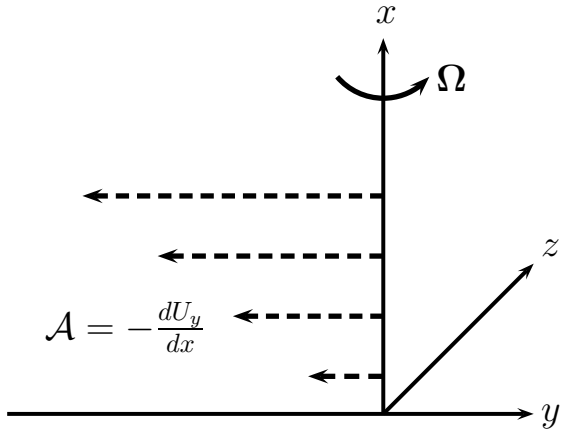}
\caption{\label{DessinEq} Sketch of the configuration in the parallel case}
\end{center}
\end{figure}
Following \cite{Kim05}, we study the effect of a large-scale shear $\UU_0 = U_0(x) \hat{j}$ on the transport properties of turbulence by writing the velocity as a sum of a shear (chosen in the $x$-direction) and fluctuations: $\uu = \UU_0 + \vv = U_0(x) \hat{j} + \vv = - x \mathcal{A} \hat{j} + \vv$. Without loss of generality, we assume $\A > 0$. In the following, we consider the configuration of Figure \ref{DessinEq} where the shear and rotation (in the $x$ direction) are parallel and  simplify notation by using ${\bf \Omega} = 2 \tilde{\bf \Omega}$. Then, the Coriolis force is simply
$\Omega [-v_z \, {\bf j} + v_y \, {\bf k}]$, where ${\bf i}$, ${\bf j}$  and ${\bf k}$  are the unit vectors associated with the Cartesian coordinates. Note that our $x-y$ coordinates are not conventional in that our $x$ and $y$ directions correspond to $y$ and $x$ in previous works (see \cite{Salhi97} for instance). Therefore, the shearing, the stream-wise and the span-wise direction correspond to the $x$, $y$ and $z$ direction, respectively. Note that the particular configuration considered here, the stationary equation for the large-scale velocity reduces to:
\EQA
- \partial_x P + F_x &=& 0 \; , \\ \nonumber
- \partial_z P + F_z - \Omega U_0(x) &=& 0 \; .
\ENA
For ${\bf F}=0$, there is no equilibrium for the large-scale flow contrary to the perpendicular case where the Coriolis force can be balanced by pressure. Thus, we assume a large-scale forcing ${\bf F}$ to maintain an equilibrium for the large-scale flow \citep[e.g.][]{Yu06}. In real physical situation, this large-scale forcing can be provided by thermal wind associated with (latitudinal) temperature gradient on large scales \citep{Kichatinov95,Salhi06}. Density/temperature fluctuations in this case can then be included as a part of the small-scale external forcing ${\bf f}$ in our formalism. Note that a similar approach was adopted by \citep{Kichatinov95} in computing the transport of momentum and particle in stars, including the polar region. A consistent treatment of fluctuating temperature requires the extension of the present work to stratified rotating sheared turbulence and will be performed in future publications.  

To calculate turbulence amplitude (or kinetic energy) and turbulent transport, we need to solve the equation for the fluctuating velocity field. To this end, we employ a quasi-linear theory \cite{Moffatt78} where the nonlinear local interactions between small scales are neglected compared to nonlocal interactions between large and small scales and obtain:  
\EQA
\label{quasi-linear}
\partial_t \vv + \UU_0 \cdot \nabla \vv + \vv \cdot \nabla \UU_0  &=& - \nabla p + \nu \nabla^2 \vv + {\bf f} - {\OOmega} \times \vv \; , \\ \nonumber
\nabla \cdot \vv &=& 0 \; ,
\ENA
where $p$ and ${\bf f}$ are respectively the small-scale components of the pressure and forcing. As noted in the introduction, this approximation, also known as the RDT \cite{Townsend76}, is justified in the case of strong shear as the latter induces a weak turbulence, leading to weak interaction between small scales which is negligible compared to the (non-local) interaction between the shear and small scales. This has in fact been confirmed by direct numerical simulations, proving the validity of the predictions of quasi-linear theory with a constant-rate shear both in the non-rotating \cite{Lee90} and rotating unforced \cite{Salhi97} turbulence and also for forced turbulence \cite{Leconte06}. Further, note that the quasi-linear analysis is also valid in the limit of rapid rotation \cite{Cambon99}.

To solve Eq. (\ref{quasi-linear}), we introduce a Fourier transform with a wave number in the $x$ direction evolving in time in order to incorporate non-perturbatively the effect of the advection by the mean shear flow \cite{Goldreich64,Townsend76,Kim05}:
\begin{equation}
\label{TFdependent}
{\bf v}({\bf x},t) = \frac{1}{(2\pi)^2} \int d^3 k \; e^{i[{ k_x(t)} x + k_y y + k_z z]}
{\bf \tilde{v}}({\bf k},t) \; ,
\end{equation}
where $k_x(t) = k_x(0) + k_y \A t$. From Eqs. (\ref{quasi-linear}) and (\ref{TFdependent}), we obtain the following set of equations for the fluctuating velocity:
\EQA
\label{System1}
\A \partial_\tau \hat{v}_x &=& -i k_y \tau \hat{p} + \hat{f}_x   \; ,\\ \nonumber
\A \partial_\tau \hat{v}_y -\A \hat{v_x} &=& -i k_y \hat{p} + \hat{f}_y + \Omega  \hat{v}_z  \; , \\ \nonumber
\A \partial_\tau \hat{v}_z &=& -i k_z \hat{p} + \hat{f}_z - \Omega \hat{v}_y \; , \\ \nonumber
0 &=& \tau \hat{v}_x + \hat{v}_y + \beta \hat{v}_z  \; .
\ENA
Here, the new variables $\hat{\vv} = \tilde{\vv} \exp[\nu (k_H^2 t + k_x^3/3k_y\A)]$ and similarly for $\hat{\bf f}$ and $\hat{p}$ have been used to absorb the diffusive term, and the time variable has been changed to $\tau = k_x(t)/k_y$. In the remainder of the paper, we solve Eq. (\ref{System1}) for the fluctuating velocity (with a vanishing velocity as initial condition). We then use these results and the correlation of the forcing (defined in \S \ref{Forcing}) to compute the turbulence intensity and transport (defined in \S \ref{SectParticle}). %We divide our study into four cases depending on the relative magnitude of the three characteristic frequencies in the problem: the diffusion rate $(\nu k^2)$, the rotation rate $\Omega$ and the shearing rate $\A$. We first consider the case of large ($\Omega \gg \A$) and weak rotation ($\Omega \ll \A$). For each of these cases, we consider the large ($\A \gg \nu k^2$) and weak ($\A \ll \nu k^2$) shear limits [THIS IS INCORRECT?? - WE CONSIDER ONLY 3 CASES??].

\subsection{Transport of angular momentum}
\label{SectionEddy}
As the large-scale velocity is in the $y$ direction, we are mostly interested in the transport in that direction. The large-scale equation for the $y$ component of velocity $\UU_0$ is given by Eq. (\ref{NSrotation}) with a supplementary term ${\bf \nabla} \cdot {\bf R}$ where ${\bf R}$ is the Reynolds stress given by:
\EQ
{\bf R} = \langle {\bf v} v_y \rangle \; .
\EN

To understand the effect of ${\bf R}$ on the transport of angular momentum, one can formally Taylor expand it with respect to the gradient of the large-scale flow:
\EQ
\label{BullShit200}
R_i = \Lambda_i U_0 - \nu_T \partial_x U_0 \delta_{i1} + \dots = \Lambda_i U_0 + \nu_T \A \delta_{i1} + \dots\; .
\EN
Here, $\Lambda_i$ and $\nu_T$ are the two turbulent transport coefficients from non-diffusive and diffusive momentum flux, respectively. Note that the first term in the expansion is due to the small-scale driving and the Coriolis force in Eq. (\ref{NSrotation}) which break the Galilean invariance \cite{Dubrulle91}.
First, $\nu_T$ is the turbulent (eddy) viscosity, which simply changes the viscosity from the molecular value $\nu$ to the effective value $\nu + \nu_T$. Note that the sign of eddy viscosity represents the direction of energy cascade, with positive (negative) value for direct (inverse) cascade. Second, the first term involving $\Lambda_i$ in equation (\ref{BullShit200}) is proportional to the rotation rate rather than the velocity gradient. This means that it does not vanish for a constant velocity field and thus permits the creation of gradient in the large-scale velocity field. This term bears some similarity with the $\alpha$ effect in dynamo theory \cite{Parker55,Steenbeck66b} and has been known as the $\Lambda$-effect \cite{Lebedinsky41,Rudiger80} or anisotropic kinetic alpha (AKA)-effect \cite{Frisch87}. Similarly to the $\alpha$ effect, this effect exists only if the small-scale flow lacks parity invariance (going from right-handed to left handed coordinates). However, in contrast to the $\alpha$ effect, the $\Lambda$ effect requires anisotropy for its existence \cite{Rudiger80,Frisch87}. 

\subsection{Particle (or heat) transport}
\label{SectParticle}
To study the influence of rotation and shear on the particle and heat transport, we have to supplement equation (\ref{NSrotation}) with an advection-diffusion equation for these quantities. We here focus on the transport of particles since a similar result also holds for the heat transport. The density of particles $N({\bf x},t)$ is governed by the following equation:
\EQ
\partial_t N + \UU \cdot \nabla N = D \nabla^2 N \; ,
\EN
where $D$ is the molecular diffusivity of particle. Note that, in the case of heat equation, $D$ should be replaced by the molecular heat conductivity $\chi$. Writing the density as the sum of a large-scale component $N_0$ and small-scale fluctuations $n$ ($N = N_0 + n$), we can express the evolution of the transport of chemicals on large scales by:
\EQ
\partial_t N_0 + \UU_0 \cdot \nabla N_0 = (D \delta_{ij} + D_T^{ij}) \partial_i \partial_j N_0 \; ,
\EN
where the turbulent diffusivity is defined as $\langle v_i n \rangle = - D_T^{ij} \partial_j N_0$. The turbulent diffusivity is computed in the following to study the effect of rotation and flow shear on turbulent transport of chemicals which can be highly anisotropic. Note that the transport of a passive scalar quantity (contrary to the angular momentum which is a vector quantity) has to be diffusive due to the fact that it is solely advected by the flow \cite{Frisch89}.

For simplicity, we assume a unit Prandtl number $D = \nu$ and apply the transformation introduced in equation (\ref{TFdependent}) to the density fluctuation $n$ to obtain the following equation: 
\EQ
\label{EquTransport}
\partial_\tau \hat{n} = \frac{(- \partial_j N_0)}{\A} \hat{v}_j \; .
\EN
Equation (\ref{EquTransport}) simply shows that the fluctuating density of particles can be obtained by integrating the fluctuating velocity in time.

\subsection{External forcing}
\label{Forcing}
As mentioned in introduction, we consider a turbulence driven by an external forcing ${\bf f}$. To calculate the turbulence amplitude and transport defined in \S \ref{SectionEddy} and \S \ref{SectParticle} (which involve quadratic functions of velocity and/or density), we prescribe this forcing to be short correlated in time (modeled by a  $\delta$-function) and homogeneous in space with power spectrum $\psi_{ij}$ in the Fourier space. Specifically, we assume:
\begin{equation}
\label{ForcingCorrel}
\langle \tilde{f}_i({\bf k_1},t_1) \tilde{f}_j({\bf k_2},t_2) \rangle = \tau_f \, (2\pi)^3 \delta({\bf k_1}+{\bf k_2}) \, \delta(t_1-t_2) \, \psi_{ij}({\bf k_2}) \; ,
\end{equation}
for $i$ and $j$ = $1$, $2$ or $3$. The angular brackets stand for an average over realizations of the forcing, and $\tau_f$ is the (short) correlation time of the forcing. Note that the $\delta$ correlation is valid as long as the correction time $\tau_f$ is the shortest time-scale in the system [i.e. $\tau_f \ll \Omega^{-1}, \A^{-1}, 1/(\nu k^2)$].

For most results that will be derived later, we assume an incompressible and isotropic forcing where the spectrum of the forcing is given by:
\EQ
\label{ForcingIsotropic}
\psi_{ij}({\bf k}) = F(k) (\delta_{ij} - k_i k_j/k^2) \; .
\EN 
It is easy to check that in the absence of rotation and shear, this forcing leads to an isotropic turbulence with intensity:
\EQ
\label{IsotropicOriginel}
\langle v_{0}^2 \rangle = \frac{2 \tau_f}{(2 \pi)^2} \int_0^\infty \frac{F(k)}{\nu} \; dk \; ,
\EN
where the subscript $0$ stands for a turbulence without shear and rotation.

%%%%%%%%%%%%%%%%%%%%%%%%%%%%%%%%%%%%%%%%%%%%%%%%%%%%%%%%%%%%%%%%%%%%%%%%%%%%%%%%%
\section{Analytical results}
\label{Pole}

To investigate turbulence property in the parallel case, we rearrange (\ref{System1}), thereby eliminating the pressure terms, obtaining the following equations for fluctuating velocity: 
\EQA
\label{System2}
\partial_\tau \Bigl[\frac{1}{\tau} \partial_\tau \bigl( \{ \gamma + \tau^2 \} \hat{v}_x \bigr) \Bigr] &+& (\bar{\Omega}^2 \tau - \bar{\Omega} \beta) \hat{v}_x = \partial_\tau \Bigl[\frac{h_1(\tau)}{\A \tau}\Bigr] - \bar{\Omega} \frac{\hat{h}_2(\tau)}{\A} \; , \\ \nonumber
\partial_\tau \hat{v}_z &=& - \frac{\beta}{\gamma} \partial_\tau \bigl[ \tau \hat{v}_x \bigr] + \frac{\OB \tau-\beta}{\gamma} \hat{v}_x + \frac{\hat{h}_2(\tau)}{\gamma \A} \; , \\ \nonumber
\hat{v}_y &=& - (\tau \hat{v}_x  + \beta \hat{v}_z) \; .
\ENA
Here, $\beta = k_z/k_y$, $\gamma = 1+\beta^2 = k_H^2 / k_y^2$ and $\OB = \Omega / \A$. To calculate the turbulence amplitude and transport, the first equation in equation (\ref{System2}) is to be solved with initial conditions: $\hat{v}_x(\tau_0) = 0$ and $\partial_\tau \hat{v}_x \vert_{\tau=\tau_0} = h_1(\tau_0) / (\gamma + \tau_0^2) \A$.  

Unfortunately, we were unable to find an exact solution of equation (\ref{System2}) in the general case. To gain a physical insight into the role of inertial waves and flow shear in turbulent transport, we  consider the two limits of strong rotation  where the effect of waves dominates shearing ($\Omega \gg \A$) and weak rotation where shearing dominates the effects of waves ($\Omega \ll \A$) in section \ref{WKBPoles} and \ref{WeakRotPoles}, respectively. Approximate solutions can be derived in these two regimes which can then be used for deriving analytic form of correlation functions for turbulence intensity and transport. However, the WKB approximation used to study the rapid rotation limit being ill-defined for some value of the parameter, we consider an exactly solvable case where all quantities can be calculated in section \ref{beta0Poles}.

\subsection{Rapid rotation limit: $\Omega \gg \A$ (and $a > 0$)}
\label{WKBPoles}
For $\vert \OB \vert \gg 1$, we seek for a WKB solution of the first equation in equation (\ref{System2}). However, since this approximation breaks for $\tau = 0$, we assume that the initial value is positive ($\tau_0 = a > 0$) to make our solution meaningful. In \S \ref{beta0Poles}, we study an exactly solvable case and show that the solution can be altered by negative initial value  ($\tau_0 = a < 0$). Assuming $\tau_0 = a > 0$, we obtain the following solutions for the three components of the velocity for $\vert \Omega / \A \vert \gg 1$:
\EQA
\nonumber
\hat{v}_x(\tau) &=& \frac{1}{\A (\gamma + \tau^2)^{3/4}} \int_{\tau_0}^\tau dt \left\{ \frac{\hat{h}_1(t)}{(\gamma+t^2)^{1/4}} \cos[v(t,\tau)] + \chi \hat{h}_2(t) (\gamma+t^2)^{1/4}  \sin[v(t,\tau)] \right\} \; , \\ \nonumber
\hat{v}_y(\tau) &=& \frac{1}{\A \gamma (\gamma + \tau^2)^{3/4}} \int_{\tau_0}^\tau dt \Bigl\{ \frac{\hat{h}_1(t)}{(\gamma+t^2)^{1/4}} \left(-\tau \cos[v(t,\tau)] + \beta \chi \sqrt{\gamma+\tau^2} \sin[v(t,\tau)] \right) \\ \label{WKBpole}
&& \qquad + \hat{h}_2(t) (\gamma+t^2)^{1/4} \left(-\chi \tau \sin[v(t,\tau)] -\beta \sqrt{\gamma+\tau^2} \cos[v(t,\tau)] \right) \Bigr\}  \; , \\ \nonumber
\hat{v}_z(\tau) &=& \frac{1}{\A \gamma (\gamma + \tau^2)^{3/4}} \int_{\tau_0}^\tau dt \Bigl\{ \frac{\hat{h}_1(t)}{(\gamma+t^2)^{1/4}} \left(- \beta \tau \cos[v(t,\tau)] - \chi \sqrt{\gamma+\tau^2} \sin[v(t,\tau)] \right) \\ \nonumber
&& \qquad + \hat{h}_2(t) (\gamma+t^2)^{1/4} \left(-\chi \beta \tau \sin[v(t,\tau)] + \sqrt{\gamma+\tau^2} \cos[v(t,\tau)] \right) \Bigr\} \; .
\ENA
Here,
\EQA
\label{Def2}
\OBN &=& \vert \OB \vert \quad , \qquad \chi = \text{sign}(\OB) \; , \\ \nonumber
r(t) &=& \sqrt{\gamma+\tau^2} -\frac{\beta \chi}{2 \OBN} \ln\left[\tau+\sqrt{\gamma+\tau^2}\right] + O\left(\frac{1}{\OBN^2}\right) \; , \\ \nonumber
v(t,\tau) &=& \OBN \left[r(t) - r(\tau)\right] \; .
\ENA
Due to the similarity between equation (\ref{WKBpole}) and the one obtained in the perpendicular case \cite{Perpendicular08}, we give only results and refer the reader to our previous paper for details of the derivation.

\subsubsection{Turbulence intensity}
In the weak shear limit ($\xi \gg 1$) where the shear is negligible, we obtain the following result for the turbulent intensity:
\EQ
\label{Bullshit600}
\langle v_x^2 \rangle = \langle v_y^2 \rangle = \langle v_z^2 \rangle = \frac{2 \tau_f}{3 (2 \pi)^2} \int d k \; \frac{F(k)}{\nu } 
=\frac{1}{3}\langle v_0^2 \rangle\; . 
\EN
Here, $\langle v_0^2 \rangle$ is the turbulence amplitude in the absence of rotation and shear [see Eq. (\ref{IsotropicOriginel})]. These results thus show that, in the large rotation limit, the turbulence intensity is isotropic and equals to the one without rotation [see Eq. (\ref{IsotropicOriginel})] for sufficiently weak shear with $\xi \gg 1$. Furthermore, in this limit of sufficiently weak shear where $(\Omega, \mathcal{D}) \gg \A$, turbulence intensity is independent of rotation since waves do not necessarily quench turbulence level. A similar result was also obtained in MHD turbulence and stratified turbulence where magnetic fields and gravity waves mainly affect transport without much effect on turbulence level \cite{Kim06,Dynamics,Stratification}. We shall show below that a strong anisotropy can be induced when shearing effect is not negligible ($\xi \ll 1$) even in the rapid rotation limit ($\Omega \gg \A$). 

In order to understand the effect of  flow shear, we now consider the strong shear limit ($\xi \ll 1$). In this limit, we obtain:
\EQA
\label{TurbulenceStrongShear}
\langle v_x^2 \rangle &=& \frac{\tau_f}{(2 \pi)^3 \A} \int d^3 k \; \sqrt{\gamma + a^2} F(k)  
\;\;\propto \; \xi \langle v_0^2 \rangle
\; , \\ \nonumber
\langle v_y^2 \rangle = \langle v_z^2 \rangle &=& \frac{\tau_f}{(2 \pi)^3 \A} \int d^3 k \; \sqrt{\gamma + a^2} F(k) \frac{-\ln\xi}{3} 
\;\;\propto \; \xi |\ln \xi|\langle v_0^2 \rangle \; ,
\ENA
to leading order in $\xi \ll 1$. Note that in the calculation of $\langle v_x^2 \rangle$, we neglected the component proportional to $a=k_x / k_y$ as it is odd in both $k_x$ and $k_y$ and thus vanishes after integration over the angular variables for an isotropic forcing. The last terms in Eq. (\ref{TurbulenceStrongShear}), expressed in terms of the turbulence amplitude in the absence of rotation and shear $\langle v_0^2 \rangle$ [see Eq. (\ref{IsotropicOriginel})], explicitly show the dependence of turbulence level on rotation and shear. That is, all the components of turbulence intensity are reduced for strong shear $\xi \ll 1$. Further, the $x$ component along shear is reduced as $\xi \propto \A^{-1}$ while the other two components as $\xi |\ln \xi|$, with an effectively weaker turbulence in the shear direction than in the perpendicular one, by a factor of $\vert \ln\xi \vert$. This shows that shear flow can induce anisotropic turbulence (unlike rotation) even when the forcing is isotropic. This result is similar to that obtained in the simulation of a Couette flow at high rotation rate \cite{Bech97} where the velocity fluctuations perpendicular to the wall exceed that in the stream-wise direction. Nevertheless, Eq. (\ref{TurbulenceStrongShear}) shows that a strong rapid rotation yet insures an isotropy in velocity fluctuations in $y-z$ directions ($\langle v_y^2 \rangle = \langle v_z^2 \rangle$). 

%This contrasts to the case of slow rotation (considered in \S \ref{WeakRotEq}) where $\langle v_y^2 \rangle$ is larger than $\langle v_z^2 \rangle$.

\subsubsection{Transport of angular momentum}
\label{EqMomentWKB}

First, in the large rotation and weak shear limit, the Reynolds stress becomes purely diffusive (with no $\Lambda$-effect) with the turbulent viscosity:
\EQ
\label{nutWKBequatorWeak}
\nu_T \sim  \frac{\pi \tau_f}{32 (2 \pi)^2 \vert \Omega \vert} \int_0^{+\infty} d k \frac{F(k)}{\nu}   \; .
\EN
This result shows that the turbulent viscosity is positive and proportional to $\Omega^{-1}$ for large $\Omega$. 

In comparison, in the strong shear limit ($\xi \ll 1$), we obtain the turbulent viscosity in the strong shear limit as:
\EQ
\label{nutWKBequator}
\nu_T = \frac{\langle v_x v_y \rangle}{\A} = - \frac{\tau_f}{(2 \pi)^3 \A^2} \int d^3 k \;   F(k)  \; .
\EN
Eq. (\ref{nutWKBequator}) shows that the turbulent viscosity is negative (as $F(k) > 0$) in the strong shear limit, in sharp contrast to the weak shear limit where $\nu_T > 0$ [see Eq. (\ref{nutWKBequatorWeak})]. Furthermore, the magnitude of $\nu_T$ is reduced by the shear ($\propto \A^{-2}$) and is independent of rotation, which should also be compared with the weak shear limit [see Eq. (\ref{nutWKBequatorWeak}) where $\nu_T \propto \Omega^{-1}$]. Therefore,  the turbulent viscosity changes from positive (for weak shear) to negative (for large shear) as the ratio of shear to dissipation increases. This result can be understood if we assume that, as in most rapidly rotating fluid, the inverse cascade is associated with the conservation of a potential vorticity \cite{Pedlovsky}. In the presence of strong shear (compared to dissipation), the potential vorticity is strictly conserved giving rise to an inverse cascade (negative viscosity). When the dissipation increases, the potential vorticity is less and less conserved and thus the inverse cascade is quenched. Our results show that there is a transition from inverse to direct cascade as the dissipation is increased. A similar behavior is also found in two-dimensional hydrodynamics (HD) where an inverse cascade can be shown to be present only for sufficient weak dissipation \cite{Kim01}.

It is important to note that the negative viscosity $\nu_T<0$ obtained here for strong rotation/strong shear ($\Omega \gg \A \gg \nu k_y^2$) signifies the amplification of shear flow as the effect of rotation favoring inverse cascade dominates shearing (generating small scales). However, the magnitude of $\nu_T$ is reduced by shear as $|\nu_T| \propto \A^{-2}$ since flow shear  inhibits the inverse cascade. This can be viewed as `self-regulation' -- that is, self-amplification of shear flow is slowed down as the latter becomes stronger.

\subsubsection{Transport of particles}
In the rapid rotation limit ($\vert \Omega \vert / \A \gg 1$), turbulent particle diffusivities can be obtained as: 
\EQA
\label{Isotropic2}
D_T^{xx} &\sim& \frac{\tau_f}{8 \pi \vert \Omega \vert} \int_0^{\infty} \frac{F(k)}{\nu}  \; dk \; , \\ \nonumber
D_T^{yy} = D_T^{zz} &\sim& \frac{\tau_f}{16 \pi \vert \Omega \vert} \int_0^{\infty} \frac{F(k)}{\nu} \; dk  \;\sim \; \frac{1}{2} D_T^{xx}\; .
\ENA
Note that in that case, the result is not sensitive to the value of the parameter $\xi$ and thus we do not distinguish between the weak and large shear limits. Eq. (\ref{Isotropic2}) shows that $D_T^{xx}$, $D_T^{yy}$ and $D_T^{zz}$ are all reduced as $\Omega^{-1}$ (with no effect of the shear) for large $\Omega$ and also that there is only a slight anisotropy in the transport of scalar: the transport in the direction of the rotation is twice larger than the one in the perpendicular direction \cite{Kichatinov94}. Interestingly, this anisotropy in the transport of particles is not present in turbulence intensity [see Eq. (\ref{Bullshit600})]. This is because waves can affect the phase between density fluctuation and velocity, not necessarily altering their amplitude. However, it is important to note that this anisotropy is only a factor of 2, much weaker than that in sheared turbulence without rotation \cite{Kim05}.

\subsection{Weak rotation limit: $\Omega \ll \A$}
\label{WeakRotPoles}
In the weak rotation limit, we expand all the quantities in powers of $\OBN = \vert \Omega / \A \vert$ as:
\EQ
X(\tau) = X_0(\tau) + \OBN X_1(\tau) + \dots \; ,
\EN
in the weak rotation limit ($\Omega \ll \A$) and calculate the turbulence intensity and transport up to first order in $\OBN$. For the sake of brevity, we here just provide the final results of the calculation. Note that in this limit, we are only interested in strong shear case ($\xi \ll 1$) since in the opposite limit where $\nu k_y^2 \gg \A \gg \Omega$, the effects of both shear and rotation simply disappear to leading order. 

\subsubsection{Turbulence intensity}
\label{BullShit1100}

In the strong shear limit ($\xi \ll 1$), we obtain the turbulence intensity up to order $\Omega$ as follows:
\EQA
\label{Vxweak}
\langle v_x^2 \rangle &=& \frac{\tau_f}{2 (2\pi)^3 \A} \int d^3 k \; (\gamma+a^2) F(k)  \left[\frac{\pi}{2 \sqrt{\gamma}} - \T(a) - \frac{a}{\gamma+a^2} \right] \; , \\ \nonumber
\langle v_y^2  \rangle &\sim& \frac{\tau_f}{(2\pi)^3 \A} \int d^3 k \; F(k) \left[(\gamma+a^2) \beta^2 \left(\frac{\pi}{2\sqrt{\gamma}} - \T(a)\right)^2 + 1 \right] \frac{\beta^2}{3 \gamma} \left(\frac{3}{2\xi}\right)^{1/3} \Gamma(1/3) \; , \\ \nonumber
\langle v_z^2  \rangle &\sim& \frac{\tau_f}{(2\pi)^3 \A} \int d^3 \; k F(k) \left[(\gamma+a^2) \beta^2 \left(\frac{\pi}{2\sqrt{\gamma}} - \T(a)\right)^2 + 1 \right] \frac{1}{3\gamma} \left(\frac{3}{2\xi}\right)^{1/3} \Gamma(1/3) \,  .
\ENA
Here, $\Gamma$ is the Gamma function. Note that the first correction (proportional to $\Omega$) vanishes and consequently, the turbulence amplitude is the same as in the case of shear without rotation \cite{Kim05}, with stronger turbulence in the direction perpendicular to the shear than in the parallel one. 
 
\subsubsection{Transport of angular momentum}
In the strong shear limit ($\xi \ll  1$), momentum flux in the azimuthal direction can be shown to be purely diffusive and given by:
\EQ
\label{LambdaEquator}
\langle v_x v_y  \rangle \sim \frac{\tau_f}{(2\pi)^3 \A} \int d^3 k \; (\gamma+a^2) F(k)  \left[-\frac{1}{2(\gamma+a^2)} + \beta^2 \left(\frac{\pi}{2\sqrt{\gamma}} - \T(a)\right)^2 \right] \; .
\EN
This recovers the eddy viscosity of sheared turbulence without rotation \cite{Kim05}, showing that its value decreases as $\propto \A^{-2}$ for strong shear. This result agrees with previous studies of non-rotating sheared turbulence \cite{Nazarenko00b} which found a Reynolds stress inversely proportional to the shear, leading to a log dependence on the distance to the wall for the large-scale shear flow. 

Alternatively, the component of the Reynolds stress involving the velocity component $v_z$ does not vanish and is odd in $\Omega$. Thus, the $\Lambda$-effect appears here in the $z$-component of the Reynolds stress $\Lambda_z$ (recall that in the perpendicular case, the $\Lambda$-effect was present only in $\langle v_x v_y \rangle$), in the form:
\EQ
\label{LambdaPole}
%\Lambda_x &\sim&  - \frac{\tau_f}{(2\pi)^3 \A^2} \int d^3 k  \frac{\Gamma(1/3)}{6 \gamma} \left(\frac{3}{2\xi}\right)^{1/3} \left\{\beta^2 \left(\frac{\pi}{2\sqrt{\gamma}} - \T(a)\right)^2 \phi_{11}({\bf k})  +  \phi_{22}({\bf k}) \right\} \; , \\ \nonumber
\Lambda_z \sim - \frac{\tau_f}{(2\pi)^3 \A^2} \int d^3 k  F(k) \frac{\Gamma(2/3)}{6 \gamma} \left(\frac{3}{2\xi}\right)^{2/3} (3 \beta^2 - 1) \left\{(\gamma+a^2) \beta^2 \left(\frac{\pi}{2\sqrt{\gamma}} - \T(a)\right)^2 + 1 \right\}\; .
\EN
Equation (\ref{LambdaPole}) shows that the sign of $\Lambda_z$ is indefinite (as both signs appear in the prefactor $(3 \beta^2 - 1)$. However, as for equation (\ref{LambdaEquator}), in the case of an isotropic forcing, the term proportional to $\beta^2$  dominates, making $\Lambda_z$ negative.  This $\Lambda$ effect appears due to the anisotropy between the stream-wise and the span-wise components of the velocity, due to the shear (alone).

\subsubsection{Transport of particles}
Up to order $\Omega$, we find the turbulent diffusivity of particles as:
\EQA
\label{TransportEquatWeak}
D_T^{xx} &\sim& \frac{\tau_f}{(2\pi)^3 \A^2} \int d^3 k  \; \gamma(\gamma+a^2) F(k) \left(\frac{\pi}{2\sqrt{\gamma}} - \T(a)\right)^2  \; , \\ \nonumber 
D_T^{zz} &\sim& \frac{\tau_f}{(2\pi)^3 \A^2} \int d^3 k \; F(k) \left[\beta^2 (\gamma+a^2) \left(\frac{\pi}{2\sqrt{\gamma}} - \T(a)\right)^2 + 1 \right] \frac{1}{3\gamma} \left(\frac{3}{2\xi}\right)^{2/3} \Gamma(2/3)  \; .
\ENA
Here again, the first correction due to rotation vanishes and one recover the result of turbulence in presence of shear alone: $D_T^{xx} \propto \A^{-2}$ and $D_T^{zz}\propto \A^{-4/3}$, with effectively faster transport in span-wise direction compared to shear direction.

\subsection{Symmetric perturbation ($\beta = 0$)}
\label{beta0Poles}
In this section, we consider a symmetric perturbation with $k_z = 0$ by assuming a forcing that is symmetric in the span-wise direction with no dependence on $z$. Note that even though $k_z=0$, $v_z$ and $v_y$ are closely linked through rotation $\Omega {\bf i}$. Details of the derivation are given in Appendix \ref{Appendixbeta0Poles}. The interest of this case is that we obtain solutions for arbitrary values of $\OBN$ and so we can look at the $\OBN \gg 1$ limit without the $a > 0$ limit. Consequently, we will show here only the results in the large rotation limit.

\subsubsection{Turbulence amplitude}
\label{AmplitudeBessel}
In the large rotation limit: $\OBN \gg 1$, we obtain the following leading order contribution of the turbulent amplitude:
\EQA
\label{AmplitudeBetaLarge}
\langle v_x^2 \rangle &=& \frac{\tau_f}{(2\pi)^3 \A} \int \, d^3 k \, F(k) \sqrt{1+a^2}   \; , \\ \nonumber
\langle v_z^2 \rangle &=& \frac{\tau_f}{(2\pi)^3 \A} \int d^3 k \, F(k) \sqrt{1+a^2}  \left( \frac{- \ln \xi}{3} \right)  \; .
\ENA
Thus, the turbulence amplitude is larger in the $y-z$ plane than the one in the shear direction by a logarithmic factor. Moreover, equation (\ref{AmplitudeBetaLarge}) shows that the turbulence amplitude does not depend on the rotation rate in the large rotation limit but is quenched by shear only. In particular, $\langle v_y^2 \rangle = \langle v_z^2 \rangle$. These results are the same as in the case where the shear and the rotation are perpendicular \cite{Perpendicular08} and thus agree with the WKB solution of Sec. \ref{WKBPoles}.

\subsubsection{Turbulent transport of momentum}
\label{MomentBessel}
In the large rotation limit ($\OBN \gg 1$), we obtain the following turbulent viscosity:
\EQ
\label{Bullshit11}
\nu_T = - \frac{\tau_f}{(2\pi)^3 \A^2}  \int \, d^3 k \, F(k) \; . 
\EN
Equation (\ref{Bullshit11}) shows that the turbulent viscosity does not depend on rotation in the large rotation limit and is obviously negative. Note that this result is the same as in the perpendicular case [see equation (\ref{nutWKBequator})] and, thus again, agrees with the WKB solution found previously.

\subsubsection{Particles transport}
\label{ParticleBessel}
In the limits of strong shear ($\xi \ll 1$) and large rotation ($\OBN \gg 1$), the transport of particles is given by: 
\EQA
\nonumber
D_T^{xx} &\sim& \frac{\tau_f}{8 \pi \vert \Omega \vert} \int_0^{\infty} \frac{F(k)}{\nu} \, dk + \frac{\pi \tau_f}{(2\pi)^3 \A \vert \Omega \vert} \int_{a < 0} \, d^3 k \sqrt{1+a^2} F(k) \; , \\ 
\label{Bullshit400}
D_T^{zz} &\sim& \frac{\tau_f}{16 \pi \vert \Omega \vert} \int_0^{\infty} \frac{F(k)}{\nu} \, dk + \frac{\pi \tau_f}{(2\pi)^3 \A \vert \Omega \vert} \int_{a < 0} \, d^3 k \sqrt{1+a^2} F(k) \; .
\ENA
The transport of particles in equation (\ref{Bullshit400}) involves two contributions, both of which scale as $\Omega^{-1}$ for rapid rotation. The first contribution comes from the integration by parts and has to be kept only because $\overline{\omega_0}$ can vanish for $a=0$ while the second comes from the stationary point in the integration (see appendix \ref{Oscill} for details). Note that the ratio of the second term to the first one is equal to $\nu k^2 / \A \sim \xi$. Consequently, in the strong shear limit ($\xi \ll 1$), the first term dominates. Thus, the transport of particles is the same as the one found with the WKB analysis (see \S \ref{WKBPoles}). 

To summarize, in this section, we solved equation (\ref{System2}) exactly for $\beta = 0$ and compared the results with the WKB analysis performed in \S \ref{WKBPoles} (which is valid only for $a > 0$). The results being the same, the conclusions reached from WKB analysis remain valid even if $a \leq 0$.

\section{Conclusion}
\label{Conclusion}
In this paper, we have performed a thorough investigation of the combined effects of shear and rotation on the structure of turbulence, by using a quasi-linear theory. We assumed an external forcing in the Navier-Stokes equation which leads to an equilibrium situation where the dissipation (whose effect is enhanced by the shear) is balanced by the injection of energy due to forcing. It is useful to recall that there are three (inverse) time-scales in the problem: the shearing rate $\A$, the rotation rate $\Omega$ and the diffusion rate $\mathcal{D}=\nu k_y^2$ where $\nu$ is the (molecular) viscosity of the fluid and $1/k_y$ is a characteristic small-scale of the forcing. The first regime of strong rotation ($\Omega \gg \A$) has been studied in the strong shear ($\A \gg \mathcal{D}$) and weak shear ($\A \ll \mathcal{D}$) limits. However, the second regime of weak rotation has been considered only in the strong shear ($\A \gg \mathcal{D}$) case, as the effects of both shear and rotation disappear in the opposite case. 

In the large rotation limit ($\vert \Omega \vert \gg \A$), we found that the results coincides with that obtained in the case where the rotation and the shear are perpendicular. Specifically, we obtained the following results:
\begin{itemize}
\item The turbulent intensity is reduced only by a strong shear  (i.e. in the case of strong rotation and strong shear) and in an anisotropic way. 
\item As the dissipation decreases (compared to the shear), there is a crossover from a positive to a negative viscosity. 
\item The transport of particle is reduced by rotation, with a slight anisotropy of a factor $2$, largely unaffected by shear.
\end{itemize}

In the opposite weak rotation limit ($\vert \Omega \vert \ll \A$), we found that the main reduction is due to the shear with an anisotropic turbulence with preferred motion and transport in the plane perpendicular to the shear. Contrary to the perpendicular case, we found here that the turbulence intensity and the particle transport are not affected by rotation.  Furthermore, we found non-diffusive flux for momentum transport (the so-called $\Lambda$-effect) which transfers energy from the fluctuating velocity field to the large-scale flow. This effect can appear even for isotropic forcing due to the fact that the shear induces an anisotropic turbulence. In this paper, the lambda effect appear on the $z$ component which has to be contrasted with the perpendicular case \cite{Perpendicular08} where the non-diffusive fluxes appeared on the $x$-component.

Table \ref{Summary} summarizes the findings of this paper together with these of paper II \cite{Perpendicular08} by highlighting the quenching of these quantities due to large shearing rate $\A$ and the rotation rate $\Omega$ (or their ratio, $\OB=\Omega / \A$). We choose to show the result in the strong shear limit ($\xi = \nu k_y^2/\A \ll 1$) as it is the proper limit to capture the effect of the shear. Furthermore, from the physical point of view, it is the meaningful limit in a vast number of systems (for example the Sun).
\renewcommand{\arraystretch}{1.8}
\begin{table}
\begin{center}
\begin{tabular}{|c|c|c|c|c|}
\hline
& \multicolumn{2}{c|}{Perpendicular \cite{Perpendicular08}} & \multicolumn{2}{c|}{Parallel} \\ \hline
& $\Omega \gg \A$ & $\Omega \ll \A$ & $\Omega \gg \A$  & $\Omega \ll \A$\\ \hline
$\langle v_x^2 \rangle$ & $\qquad \A^{-1} \qquad$ & $\A^{-1} \left[1+ C \OB\right]$ & $\qquad \A^{-1} \qquad$ & $\qquad \A^{-1} \qquad $ \\
$\langle v_y^2 \rangle \sim \langle v_z^2 \rangle$ & $ \A^{-1} \vert \ln \xi \vert$  & $\A^{-2/3} \left[1+ C  \OB \vert \ln \xi \vert\right]$ & $\A^{-1}\vert \ln \xi \vert$ & $\A^{-2/3}$ \\ \hline
$ \nu_T $ & $-\A^{-2}$ & $\A^{-2} $ & $-\A^{-2}$ & $\A^{-2}$ \\
$ \Lambda_x $ & $0$ & $\A^{-2} \vert \ln \xi \vert$ & $0$ & $0$ \\
$ \Lambda_z $ & $0$ & $0$ & $0$ & $\A^{-4/3}$ \\ \hline
$D_T^{xx}$ & $\Omega^{-1}$ &  $\A^{-2} \left[1+ C \OB \vert \ln \xi \vert\right]$ & $\Omega^{-1}$ & $\A^{-2}$ \\
$D_T^{yy} \sim D_T^{zz}$ & $\Omega^{-1}$ & $\A^{-4/3} \left[1+ C \OB \vert \ln \xi \vert\right]$ & $\Omega^{-1}$ & $\A^{-4/3}$ \\ \hline
\end{tabular}
\end{center}
\caption{\label{Summary} Summary of our results obtained for the perpendicular \cite{Perpendicular08} and parallel cases in the strong shear limit ($\xi = \nu k_y^2/\A \ll 1$) and for an isotropic forcing. In the perpendicular case, the rotation is in the $z$ direction whereas it is in the $x$ direction in the parallel case. In both cases, the shear is in the $x$ direction. The $C$ symbol stands for an additional constant of order $1$.}
\end{table}

These results can have significant implications for astrophysical and geophysical systems. For instance, the $\Lambda$-effect and/or negative viscosity can provide a mechanism for the generation of ubiquitous large-scale shear flows, which are often observed in these objects. Furthermore, the anisotropic mixing of scalars should be taken into account in understanding the surface depletion of light elements in stars \cite{Pinsonneault97}. Finally, we note that numerical confirmation of our prediction and the extension of our work to stratified rotating sheared turbulence with/without magnetic fields remain challenging important problems, and will be addressed in future publications.

\begin{acknowledgments}
This work was supported by U.K. PPARC Grant No. PP/B501512/1 and STFC Grant No. ST/F501796/1.
\end{acknowledgments}

\begin{appendix}
\section{Symmetric perturbation ($\beta = 0$)}
\label{Appendixbeta0Poles}
In this section, we consider a symmetric perturbation with $k_z = 0$ by assuming a forcing that is symmetric in the span-wise direction with no dependence on $z$. Note that even though $k_z=0$, $v_z$ and $v_y$ are closely linked through rotation $\Omega {\hat x}$. For $\beta = k_z / k_y = 0$, the homogeneous part of the first equation in (\ref{System2}) becomes:
\EQ
\partial_\tau \Bigl[\frac{1}{\tau} \partial_\tau \bigl( \{1 + \tau^2 \} \hat{v}_x \bigr) \Bigr] + \bar{\Omega}^2 \tau \hat{v}_x = 0\; .
\EN
Solutions of the homogeneous problem are thus Bessel functions. Using the method of variation of parameters, we can then express the general solution of the first equation to (\ref{System2}) as:
\EQA
\label{VxBessel}
\hat{v}_x(\tau) &=& \frac{\pi \OBN }{2\sqrt{1+\tau^2}} \int_{\tau_0}^\tau \, dt  \left[ \frac{h_1(t)}{\A} \L_{01}(t,\tau) + \frac{h_2(t)}{\A} \sqrt{1+t^2}  \L_{11}(t,\tau) \right] \; . \\ \nonumber
\ENA
Here again, $\Omega_0 = \vert \OB \vert$, $\chi = \text{sign}(\OB)$; and $\L_{np}$ are defined by:
\EQ
\label{Bullshit700}
\L_{np}(t,\tau) = Y_n[\OBN \sqrt{1+t^2}] J_p[\OBN \sqrt{1+\tau^2}] - J_n[\OBN \sqrt{1+t^2}] Y_p[\OBN \sqrt{1+\tau^2}] \; .
\EN
The second equation of system (\ref{System2}) can then be used to obtain the other components of the velocity in the form:
\EQA
\label{TurbHoriz2}
\label{VzBessel}
\hat{v}_z(\tau) &=& \frac{\pi \OBN}{2} \int_{\tau_0}^\tau \, dt \left[ \frac{h_1(t)}{\A}  \chi  \L_{00}(t,\tau) - \frac{h_2(t)}{\A} \sqrt{1+t^2}  \L_{10}(t,\tau) \right] \; , 
\ENA
and a similar expression for $\hat{v}_y(\tau)$. We can now use equations (\ref{VxBessel}) and (\ref{VzBessel}) to calculate turbulence amplitude (\S \ref{AppendixAmplitudeBessel}) and transport (\S \ref{AppendixMomentBessel} and \S \ref{AppendixParticleBessel}). Note that equations (\ref{VxBessel}) and (\ref{VzBessel}) are exact solutions valid for all values of $\OBN$. 

\subsection{Turbulence amplitude}
\label{AppendixAmplitudeBessel}
From equations (\ref{VxBessel}) and (\ref{VzBessel}), we can easily obtain the turbulence amplitude as:
\EQA
\label{Turbulencebeta}
\langle v_x^2 \rangle = \frac{\tau_f \pi^2 \OBN^2}{4 (2\pi)^3 \A} \int d^3 k F(k) (1+a^2) \left[X_{1}({\bf k}) + X_{2}({\bf k}) \right] \; , \\ \nonumber
\langle v_z^2 \rangle = \frac{\tau_f \pi^2 \OBN^2}{4 (2\pi)^3 \A} \int d^3 k F(k) (1+a^2) \left[X_{3}({\bf k}) + X_{4}({\bf k}) \right] \; . 
\ENA
Here, for simplicity, we considered only an isotropic forcing, given by equation (\ref{ForcingIsotropic}), and defined the following integrals:
\EQA
\label{IntegralsBessel}
X_{1}({\bf k}) &=&  \int_a^{+\infty} \frac{e^{-2\xi \left[Q(\tau)-Q(a)\right]}}{1+\tau^2} \left[\L_{01}(a,\tau)\right]^2 d\tau \; , \\ \nonumber
X_{2}({\bf k}) &=&  \int_a^{+\infty} \frac{e^{-2\xi \left[Q(\tau)-Q(a)\right]}}{1+\tau^2} \left[\L_{11}(a,\tau)\right]^2 d\tau \; , \\ \nonumber
X_{3}({\bf k}) &=&  \int_a^{+\infty} e^{-2\xi \left[Q(\tau)-Q(a)\right]} \left[\L_{00}(a,\tau)\right]^2 d\tau \; , \\ \nonumber
X_{4}({\bf k}) &=&  \int_a^{+\infty} e^{-2\xi \left[Q(\tau)-Q(a)\right]} \left[\L_{10}(a,\tau)\right]^2 d\tau \; . 
\ENA
Here, $\L_{np}$'s are given by equation (\ref{Bullshit700}). We now consider the strong shear limit: $\xi = \nu k_y^2 / A \ll 1$. As both Bessel functions becomes as $(1+\tau^2)^{-1/4}$ (up to a trigonometric functions) for large $\tau$, the first two integrals converge as $\xi \rightarrow 0$. Thus, it is sufficient to put $\xi = 0$ in $X_1$ and $X_2$ in equation (\ref{IntegralsBessel}) to obtain the leading order behavior for $\xi \ll 1$. In comparison, the integrand of $X_3$ and $X_4$ behaves as $1/\tau$ for $\tau \gg 1$, giving a contribution of order $\ln\xi$ to leading order.

We now examine the turbulence amplitude in the large rotation limit: $\OBN \gg 1$. To do so, we use the asymptotic behavior of the integrals (\ref{IntegralsBessel}) derived in appendix \ref{NonOscillLarge}. Using equations (\ref{AmplitudeLargeRotation}) and (\ref{AssymptotLog}) in equation(\ref{Turbulencebeta}), we obtain the leading order contribution of the turbulent amplitude given by Eq. (\ref{AmplitudeBetaLarge}) in the main text.

\subsection{Turbulent transport of momentum}
\label{AppendixMomentBessel}
We now calculate the turbulent viscosity $\nu_T$ defined by $\langle v_x v_y \rangle = -\nu_T \partial_x U_0 = \nu_T \A$. From equations (\ref{VxBessel}) and (\ref{VzBessel}), we can derive the Reynolds stress in the case of an isotropic forcing:
\EQ
\label{Momentbeta}
\langle v_x v_y \rangle = - \frac{\tau_f \pi^2 \OBN^2}{4 (2\pi)^3 \A} \int d^3 k F(k) (1+a^2) \left[X_{5}({\bf k}) + X_{6}({\bf k}) \right] \; , 
\EN
where,
\EQA
\label{IntegralsBessel2}
X_{5}({\bf k}) &=&  \int_a^{+\infty} \frac{\tau \; e^{-2\xi \left[Q(\tau)-Q(a)\right]}}{1+\tau^2} \left[\L_{01}(t,\tau)\right]^2 d\tau \; , \\ \nonumber
X_{6}({\bf k}) &=&  \int_a^{+\infty} \frac{\tau \; e^{-2\xi \left[Q(\tau)-Q(a)\right]}}{1+\tau^2} \left[\L_{11}(a,\tau)\right]^2 d\tau \; .
\ENA
Here, $\L_{np}$'s are again given by equation (\ref{Bullshit700}). Note that the expression for the transport of angular momentum [equation (\ref{Momentbeta})] is the same as that of $\langle v_x^2 \rangle$ [equation (\ref{Turbulencebeta})] except for the multiplicative factor of $-\tau$. This is simply because, for $\beta = 0$, the incompressibility condition imposes $\hat{v}_y = -\tau \hat{v}_x$. By using the asymptotic behavior of Bessel functions for large argument, we see that the two integrals $X_5$ and $X_6$ in equation (\ref{IntegralsBessel2}
)  can be evaluated in the strong shear limit by just putting $\xi =0$. Consequently, the turbulent viscosity is of order $\A^{-2}$ for any value of $\OB$.

In the large rotation limit ($\OBN \gg 1$), we can estimate the integrals (\ref{IntegralsBessel2}) and obtain the turbulent viscosity given by Eq. (\ref{Bullshit11}) in the main text.

\subsection{Particles transport}
\label{AppendixParticleBessel}

The fluctuating concentration of particles can be obtained by integration of the fluctuating velocities (\ref{VxBessel}) and (\ref{VzBessel}) [see equation (\ref{EquTransport})]. Then, the diagonal part of turbulent diffusivity can be obtained as:
\EQA
\label{Particlebeta}
D_T^{xx} &=& \frac{\tau_f  \pi^2 \OBN^2 }{4 (2\pi)^3 \A^2} \int \, d^3 k \, (1+a^2) F(k) \left[P_{1}({\bf k}) + P_{2}({\bf k}) \right] \; , \\ \nonumber
D_T^{zz} &=& \frac{\tau_f  \pi^2 \OBN^2 }{4 (2\pi)^3 \A^2}  \int \, d^3 k \, (1+a^2) F(k) \left[P_{3}({\bf k}) + P_{4}({\bf k}) \right]  \; .
\ENA
Here, we defined integrals $P_i$ which all have the following form:
\EQ
\label{IntOscill}
P_i({\bf k}) = \int_a^{+\infty} d\tau e^{-2\xi \left[Q(\tau)-Q(a)\right]} F_i(\tau) \int_a^\tau F_i(t) \, dt \; ,
\EN
for $i=1$ to $4$. The functions $F_i(\tau)$'s are defined by:
\EQA
F_{1} =   \frac{\L_{01}(a,\tau)}{\sqrt{1+\tau^2}} \; &,& \qquad F_{2} =  \frac{\L_{11}(a,\tau)}{\sqrt{1+\tau^2}} \; , \\ \nonumber
F_{3} = \L_{00}(a,\tau)  \; &,& \qquad  F_{4} =   -  \L_{10}(a,\tau)  \; . 
\ENA  

In the large rotation limit ($\OB \gg 1$), the $F_i$'s are oscillating functions. Thus, to evaluate integrals (\ref{IntOscill}) in the strong shear limit ($\xi \ll 1$), we can not simply put $\xi =0$ in equation (\ref{IntOscill}) as is explained in the appendix \ref{Oscill}. A careful analysis (see appendix \ref{Oscill}) then gives us Eq. (\ref{Bullshit400}) of the main text in the limits of strong shear ($\xi \ll 1$) and large rotation ($\OBN \gg 1$).

\section{Asymptotic expansion of integrals}
In \S \ref{beta0Poles}, we took a large shear limit($\xi \ll 1$) and obtain equation (\ref{Turbulencebeta}) for the turbulence intensity, equation (\ref{Momentbeta}) for the transport of angular momentum, and equation (\ref{Particlebeta}) for the transport of particles in terms of integrals involving Bessel functions of an argument depending on the rotation. We  here derive asymptotic behavior of these integrals to simplify our results.

\subsection{Non Oscillating integrands}
For non oscillating integrands, it is sufficient to put $\xi = 0$ in the integrals to find the large shear limit (the resulting integral converges as $\xi \rightarrow 0$). Here, we provide asymptotic behavior of the following integrals for small or large $\OBN$:
\EQA
\label{IntnonOscill}
X_{1}({\bf k}) &=&  \int_a^{+\infty} \frac{1}{1+\tau^2} \left[\L_{0,1}(t,\tau)\right]^2 d\tau \; , \\ \nonumber
X_{2}({\bf k}) &=&  \int_a^{+\infty} \frac{1}{1+\tau^2} \left[\L_{11}(a,\tau)\right]^2 d\tau \; , \\ \nonumber
X_{5}({\bf k}) &=&  \int_a^{+\infty} \frac{\tau}{1+\tau^2} \left[\L_{01}(t,\tau)\right]^2 d\tau \; , \\ \nonumber
X_{6}({\bf k}) &=&  \int_a^{+\infty} \frac{\tau}{1+\tau^2} \left[\L_{11}(a,\tau)\right]^2 d\tau \; .
\ENA
Here $\L_{np}$'s are given by equation (\ref{Bullshit700})

\subsubsection{Small rotation limit ($\OBN \ll 1$)}
To calculate $X_1$ and $X_5$, one can use the asymptotic expansion of the Bessel functions and readily obtain:
\EQA
X_5 &\sim& \frac{4}{\pi^2 \OBN^2} \int_a^{\infty} \frac{d \tau}{(1 + \tau^2)^2} = \frac{2}{\pi^2 \OBN^2} \Bigl[\bigl(\frac{\pi}{2} - \arctan(a)\bigr) - \frac{a}{1+a^2} \Bigr] \; , \\ \nonumber 
X_5 &\sim& \frac{4}{\pi^2 \OBN^2} \int_a^{\infty} \frac{\tau d\tau}{(\gamma + \tau^2)^2} = \frac{2}{\pi^2 \OBN^2 (1+a^2)}  \; . 
\ENA
If we apply the same strategy to the calculations of $X_2$ and $X_6$, the resulting expression would not be integrable so we have to calculate it otherwise:
\EQA
X_3 &\sim& \frac{2}{\pi \OBN \sqrt{1+a^2}} \int_{a}^{\infty} \frac{J_1^2(\OBN \sqrt{ 1+\tau^2})}{1+\tau^2} \, d\tau = \frac{2}{\pi \sqrt{1+a^2}} \int_{\OBN a}^{\infty} \frac{J_1^2(\sqrt{\OBN^2+x^2})}{\OBN^2+x^2} \, d\tau \\ \nonumber
&& \qquad \sim \frac{2}{\pi  \sqrt{1+a^2}} \int_{0}^{\infty} \frac{J_1^2(x)}{x^2} \sim \frac{8}{3 \pi^2 \sqrt{1+a^2}}\; , \\ \nonumber
X_6 &\sim& \frac{2}{\pi \OBN \sqrt{1+a^2}} \int_{a}^{\infty} \frac{\tau J_1^2(\OBN \sqrt{ 1+\tau^2})}{1+\tau^2} \, d\tau \sim \frac{2}{\pi \OBN \sqrt{1+a^2}} \int_{0}^{\infty} \frac{J_1^2(x)}{x} \sim \frac{1}{\pi \OBN \sqrt{1+a^2}}  \; . 
\ENA

\subsubsection{Large rotation limit ($\OBN \gg 1$)}
\label{NonOscillLarge}
Using the Bessel asymptotic behavior for large argument, we obtain the following formula for the first integral:
\EQA
X_{1} &\sim&  \frac{4}{\pi^2 \OBN^2 \sqrt{1+a^2}} \int_a^{+\infty} \frac{ \cos^2\left[\OBN\{\sqrt{1+a^2}-\sqrt{1+\tau^2}\}\right]}{(1+\tau^2)^{3/2}} \, d\tau \\ \nonumber
&\sim& \frac{2}{\pi^2 \OBN^2 \sqrt{1+a^2}} \int_a^{+\infty} \frac{1}{(1+\tau^2)^{3/2}} = \frac{2}{\pi^2 \OBN^2 \sqrt{1+a^2}} \bigl(1-\frac{a}{\sqrt{1+a^2}}\Bigr) \; ,
\ENA
ans similarly for the other three integrals. Finally, we obtain the following asymptotic behavior for the four integrals (\ref{IntnonOscill}):
\EQA
\label{AmplitudeLargeRotation}
X_1 &\sim& X_2 \sim \frac{2}{\pi^2 \OBN^2 \sqrt{1+a^2}} \bigl(1-\frac{a}{\sqrt{1+a^2}}\Bigr) \; , \\ \nonumber
X_5 &\sim& X_6 \sim \frac{2}{\pi^2 \OBN^2} \frac{1}{1+a^2}  \; .
\ENA

\subsection{Logarithmic divergence}
\label{Logarithmic}
As noticed in \S \ref{AppendixAmplitudeBessel}, there is a logarithmic divergence arising in the calculation of $X_3$ and $X_4$. We here calculate this divergence in the case of a fast oscillation. Following \cite{Kim05}, we change the integration variable from $\tau$ to $y=2 \xi \tau^3 / 3$, replace the Bessel function by the expression valid for large argument ($\xi \ll 1$), and then obtain the following, to leading order in $\xi$:
\EQA
\label{Bullshit10}
X_3({\bf k_1}) &=& \frac{2}{\pi \OBN} \int_{\xi a^3}^{\infty}  dy \frac{e^{-y} dy}{(3 y)^{2/3} (2\xi)^{1/3} \sqrt{1+\bigl(\frac{3 y}{2 \xi}\bigr)^{2/3}}} \times \\ \nonumber 
&&  \Bigl\{\cos\bigl[ \OBN \sqrt{1+\bigl(\frac{3 y}{2 \xi}\bigr)^{2/3}}-\frac{\pi}{4} \bigr]  Y_0[w(a)] - \sin\bigl[ \OBN \sqrt{1+\bigl(\frac{3 y}{2 \xi}\bigr)^{2/3}}-\frac{\pi}{4} \bigr] J_0[w(a)] \Bigr\}^2 \,  \; . 
\ENA
We see that as $\xi$ tends to zero, the integrand in equation (\ref{Bullshit10}) becomes proportional to $1/y$, giving a contribution of the order $\ln\xi$. 

In the large rotation limit ($\OBN \gg 1$), we replace the Bessel functions by their asymptotic behavior to obtain:
\EQA
\label{AssymptotLog}
X_3 &\sim& \frac{4}{\pi^2 \OBN^2 \sqrt{1+a^2}} \int_{\xi a^3}^{\infty}  \frac{e^{-y} dy}{(3 y)^{2/3} (2\xi)^{1/3}} \frac{\sin^2\Bigl[\OBN \bigl( \sqrt{1+a^2} -  \sqrt{1+\bigl(\frac{y}{\xi}\bigr)^{2/3}}\bigr)\Bigr]}{\sqrt{1+\bigl(\frac{3y}{2\xi}\bigr)^{2/3}}}  \;  \\ \nonumber
&\sim& \frac{2}{\pi^2 \OBN^2 \sqrt{1+a^2}} \int_{\xi a^3}^{\infty}  \frac{1}{3} \frac{e^{-y} dy}{\sqrt{\left(\frac{2 \xi y^2}{3}\right)^{2/3}+y^2}} \sim \frac{2}{\pi^2 \OBN^2 \sqrt{1+a^2}} \frac{-\ln\xi}{3}\; ,
\ENA
to leading order in $\xi \ll 1$. Following the same analysis, we find the same asymptotic behavior for $X_4$.

\subsection{Oscillating integrands}
\label{Oscill}
The calculation of the transport of particles involves the computation of double integrals of the type:
\EQ
P = \int_a^{+\infty} d\tau e^{-2\xi \left[Q(\tau)-Q(a)\right]} F(\tau) \int_a^\tau F(t) \, dt \; ,
\EN
where the functions $F$ contains an oscillating functions. We  here derive the asymptotic behavior of this integral with $F(t) = f(t) \cos[\OBN \phi(t)]$ and the phase given by
$\phi(t) = \sqrt{1+a^2} - \sqrt{1+t^2}$. The difficulty associated with the calculation of such integral is the presence of a point of stationary phase $t=0$ where the integral cannot be done with an integration by part.

For $a > 0$, the point of stationary phase is never reached and then, the first integral can be approximated, for $\OBN \gg 1$,  as:
\EQ
I(\tau) \equiv \int_a^\tau F(t) \, dt \sim - \frac{\sqrt{1+\tau^2} f(\tau)}{\OBN \tau} \sin[\OBN \phi(\tau)] \; .  
\EN 
Using this approximation, $P$ can be computed with the following result:
\EQ
\label{IntegralUsuel}
P \sim \frac{f(a)^2 (1+a^2)}{4[\xi (1+a^2)^2 + \Omega_0^2 a^2]} \; .
\EN
Note that the result is the same as in the perpendicular case where the integral defining the transport of particles does not involve any stationary point.

\begin{figure}
\begin{center}
\includegraphics[scale=0.7,clip]{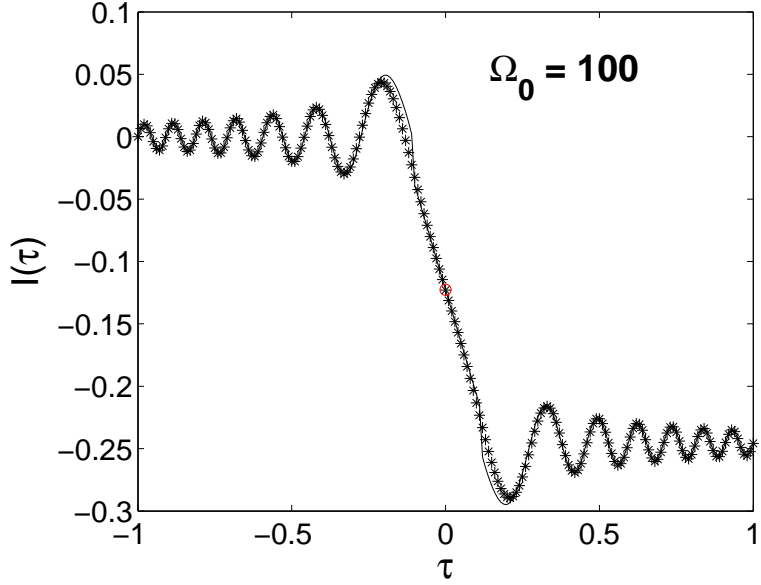}
\caption{\label{DessinApproximI} Graph of the function $I(\tau)$ with our approximation (\ref{ApproximI}). The parameters are $a = -1$ and $\OBN = 100$.}
\end{center}
\end{figure}

For $a < 0$, the behavior of the integral $I(\tau)$ is affected by the stationary point in the vicinity of $\tau = 0$. We can however find an approximation as: 
\EQ
\label{ApproximI}
I(\tau) \sim 
\begin{cases}
- \frac{\sqrt{1+\tau^2} f(\tau)}{\OBN \tau} \sin[\OBN \phi(\tau)]   & \text{if} \; \tau \leq -\frac{1}{\sqrt{\Omega_0}} \; , \\
I_0 + c \tau & \text{if} \; \vert \tau \vert < \frac{1}{\sqrt{\Omega_0}} \; , \\
2 I_0 - \frac{\sqrt{1+\tau^2} f(\tau)}{\OBN \tau} \sin[\OBN \phi(\tau)]   & \text{if} \; \tau \geq \frac{1}{\sqrt{\Omega_0}}  \; . \\
\end{cases}
\EN 
Here, $I_0  = \sqrt{\pi / 2\OBN} f(0) \cos\left[\OBN \phi(0) - \pi / 4\right]$ is the value given by the stationary point and $c=f(0) \cos\left[\OBN \phi(0)\right]$ is obtained by Taylor expanding $I$ in the vicinity of $\tau = 0$. Figure \ref{DessinApproximI} shows the numerical computation of the integral compared to the approximation (\ref{ApproximI}) and shows an excellent agreement. Using equation  (\ref{ApproximI}), we obtain $P$ as:
\EQ
P \sim \frac{f(a)^2 (1+a^2)}{4[\xi (1+a^2)^2 + \Omega_0^2 a^2]} + 2 I_0^2 \; .
\EN
The first contribution comes from the integration by part (and as the result is odd in $\tau$, the contributions from $- 1/\sqrt{\Omega_0}$ and $1/\sqrt{\Omega_0}$ cancel out). The second contribution (of order $\OBN^{-1}$) comes from the stationary point. Both contributions have to be kept as the first one can be important if $\vert \omega_0 a \vert \ll 1$.

For $a=0$, the stationary point counts twice as less, so the approximation becomes:
\EQ
I(\tau) \sim 
\begin{cases}
c \tau & \text{if} \;  0 \leq  \tau  < \frac{1}{\sqrt{\Omega_0}} \\
I_0 - \frac{\sqrt{1+\tau^2} f(\tau)}{\OBN \tau} \sin[\OBN \phi(\tau)]   & \text{if} \; \tau \geq \frac{1}{\sqrt{\Omega_0}}  \\
\end{cases} 
\EN 
In that case, the contribution from the stationary point cancels out as $I(0) = 0$. Therefore, for $a=0$, the only contribution comes from the end point of the integration and is the same as for $a > 0$ [see equation (\ref{IntegralUsuel})].

Performing the same procedure when $F(t) = f(t) \sin[\OBN \phi(t)]$, we obtain the following result:
\EQA
P &\sim& - \frac{f(a)^2 (1+a^2)}{4[\xi (1+a^2)^2 + \Omega_0^2 a^2]} + \frac{f(a)^2 (1+a^2)}{[4\xi (1+a^2)^2 + \Omega_0^2 a^2]}\\
&& \qquad \left( \quad + \quad  \frac{\pi}{\OBN} f(0)^2 \sin^2\left[\OBN \phi(0) - \pi / 4\right] \quad \right) \; ,
\ENA
the second line being present only if $a < 0$ (\ie when the point of stationary phase is reached).

\end{appendix}

\bibliographystyle{apsrev}
\bibliography{../../../Biblio/Bib_sun,../../../Biblio/Bib_shear,../../../Biblio/Bib_Geo,../../../Biblio/Bib_turbu,../../../Biblio/Bib_maths,../../../Biblio/Bib_dynamo,Bib_RotShear}

\end{document}